# New Lithium- and Diamine-Intercalated Superconductors

# $Li_x(C_nH_{2n+4}N_2)_yMoSe_2$ ($n$ = 2, 6)

Kazuki Sato, Takashi Noji, Takehiro Hatakeda, Takayuki Kawamata,
Masatsune Kato, and Yoji Koike

*Department of Applied Physics, Tohoku University, 6-6-05 Aoba, Aramaki, Aoba-ku,
Sendai 980-8579, Japan*

We have succeeded in synthesizing new intercalation superconductors $Li_x(C_2H_8N_2)_yMoSe_2$ and $Li_x(C_6H_{16}N_2)_yMoSe_2$ with $T_c$ = 4.2 and 3.8 – 6.0 K, respectively, via the co-intercalation of lithium and ethylenediamine or hexamethylenediamine into semiconducting 2H-MoSe$_2$. It has been found that the $T_c$ values are related not to the interlayer spacing between MoSe$_2$ layers so much but to the electronic density of states (EDOS) at the Fermi level. Moreover, only Li-intercalated Li$_x$MoSe$_2$ with a small interlayer spacing has been found to be non-superconducting. Accordingly, it has been concluded that not only a sufficient amount of EDOS at the Fermi level due to the charge transfer from intercalated Li to MoSe$_2$ layers but also the enhancement of the two-dimensionality of the crystal structure and/or electronic structure due to the expansion of the interlayer spacing between MoSe$_2$ layers is necessary for the appearance of superconductivity in MoSe$_2$-based intercalation superconductors. The pairing mechanism and the analogy to the superconductivity in the electric double-layer transistors of 2H-Mo$X_2$ ($X$ = S, Se, Te) are discussed.

**1. Introduction**

Transition-metal dichalcogenides $MX_2$ ($M$ = transition metal, $X$ = S, Se, Te) are typical two-dimensional layered compounds.[1] It is known that a variety of metal atoms and organic molecules are intercalated between weakly bound $MX_2$ layers, leading to marked changes in the physical properties.[2] In particular, electron doping due to the intercalation of metal atoms is an effective way to induce superconductivity in semiconducting $MX_2$. In semiconducting 2H-Mo$X_2$ ($X$ = S, Se), the intercalation of metal atoms using the liquid ammonia technique has produced several superconducting materials.[3-5] Recently, Miao et al.[6] have reported that superconductivity with the superconducting transition temperature $T_c$ = 6.0 – 7.5 K is induced in $A_x(NH_3)_yMoSe_2$

($A$ = Li, Na, K, Sr) via the co-intercalation of an alkali metal or alkaline-earth metal and ammonia into 2H-MoSe$_2$. Interestingly, the $T_c$ value tends to increase with increasing $c$-axis length owing to the increase in the ionic radius of $A$. This tendency has also been seen in $A_x$(NH$_3$)$_y$MoS$_2$ ($A$ = Li, Na, K, Rb, Cs, Ca, Sr, Ba, Yb).[3,4] These results suggest that the enhancement of the two-dimensionality due to the increase in the $c$-axis length leads to the increase in $T_c$. Therefore, it is interesting to determine whether $T_c$ increases with a further increase in the $c$-axis length. Moreover, interesting results have been reported that superconductivity is induced by the electrostatic carrier doping in thin flakes of 2H-Mo$X_2$ ($X$ = S, Se, Te) using the electric double-layer transistor (EDLT) technique.[7,8] The maximum $T_c$ values of 2H-Mo$X_2$ ($X$ = S, Se, Te) reach 10.8, 7.1, and 2.8 K, respectively, and the $T_c$ value is controlled by the electron density accumulated in the surface layer of thin flakes.

Very recently, we have reported that both ethylenediamine (EDA) C$_2$H$_8$N$_2$ and hexamethylenediamine (HMDA) C$_6$H$_{16}$N$_2$ are intercalated into semimetallic 1T-TiSe$_2$ together with lithium.[9] It has been found that $T_c$ is 4.2 K for Li$_x$(C$_2$H$_8$N$_2$)$_y$TiSe$_2$ with a $c$-axis length of 11.6 Å and also for Li$_x$(C$_6$H$_{16}$N$_2$)$_y$TiSe$_2$ with a $c$-axis length of 14.1 Å. Since 2H-MoSe$_2$ has double Se layers weakly bound by the van der Waals force as well as 1T-TiSe$_2$, as shown in Fig. 1(a), the co-intercalation of lithium and EDA or HMDA into 2H-MoSe$_2$ and the resultant appearance of superconductivity are expected.

In this study, we have attempted the co-intercalation of lithium and EDA or HMDA into 2H-MoSe$_2$ and succeeded in the synthesis of new intercalation superconductors of Li$_x$(C$_2$H$_8$N$_2$)$_y$MoSe$_2$ and Li$_x$(C$_6$H$_{16}$N$_2$)$_y$MoSe$_2$ with $T_c$ = 4.2 and 3.8 – 6.0 K, respectively, and very long $c$-axis lengths. We also measured the magnetic susceptibility $\chi$ in the normal state, and the relationship between $T_c$ and the interlayer spacing between MoSe$_2$ layers or the electronic density of states (EDOS) at the Fermi level is discussed. Moreover, effects of post-annealing on the crystal structure and superconductivity in Li$_x$(C$_2$H$_8$N$_2$)$_y$MoSe$_2$ have also been investigated.

2. Experimental

Polycrystalline host samples of 2H-MoSe$_2$ were prepared by the solid-state reaction method. Molybdenum powder and selenium grains were weighted in a molar ratio of Mo : Se = 1 : 2, mixed thoroughly, and pressed into pellets. The pellets were sealed in an evacuated quartz tube and heated at 800°C for 150 h. The obtained pellets of 2H-MoSe$_2$ were pulverized into powder to be used for the intercalation. Both Li- and EDA- or HMDA-intercalated samples of Li$_x$(C$_2$H$_8$N$_2$)$_y$MoSe$_2$ or Li$_x$(C$_6$H$_{16}$N$_2$)$_y$MoSe$_2$ were prepared as follows.[9-12] An appropriate amount of the

2H-MoSe$_2$ powder was placed in a beaker filled with 0.2 M solution of pure Li metal dissolved in EDA or HMDA. The amount of 2H-MoSe$_2$ was in a molar ratio of Li : MoSe$_2$ = 0.25 − 1 : 1. The reaction was carried out at 50˚C for 2 weeks. The product in the residual EDA was washed with fresh EDA. On the other hand, the separation of the product in the residual HMDA was easily performed by the solidification of residual HMDA at the top cap of the beaker, keeping the temperature of the top cap of the beaker below the melting point of HMDA (42˚C). All the processes were carried out in an argon-filled glove box. To investigate the change in the physical properties caused by the deintercalation of EDA, as-intercalated samples were placed in a glass tube, evacuated using an oil-rotary pump, and annealed at 250 ˚C for 20 h. Note that the $x$ values in the intercalated samples of Li$_x$(C$_2$H$_8$N$_2$)$_y$MoSe$_2$ and Li$_x$(C$_6$H$_{16}$N$_2$)$_y$MoSe$_2$ are nominal ones based on the assumption that all Li$^+$ ions in the solution are intercalated into 2H-MoSe$_2$.[12]

Both the host sample of 2H-MoSe$_2$ and the intercalated samples were characterized by powder x-ray diffraction using CuK$_\alpha$ radiation. For the intercalated samples, an airtight sample holder was used. The diffraction patterns were analyzed using RIETAN-FP.[13] $\chi$ was measured using a superconducting quantum interference device (SQUID) magnetometer (Quantum Design, MPMS). Measurements of the electrical resistivity $\rho$ were also carried out by the standard dc four-probe method. For the $\rho$ measurements, powder as-intercalated samples were pressed into pellets at room temperature without heat treatment. Thermogravimetric (TG) measurements were performed in a flow of argon gas using a commercial analyzer (SII Nano Technology Inc., TG/DTA7300).

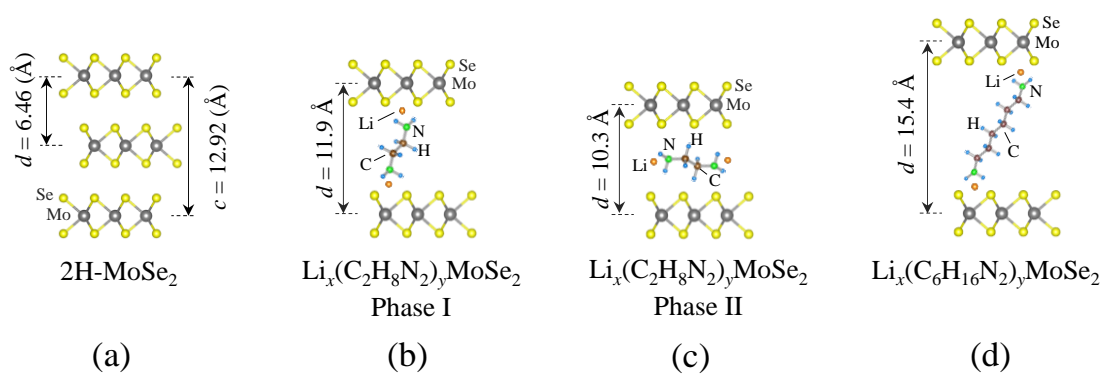

Fig. 1. Schematic views of crystal structures of (a) 2H-MoSe$_2$, (b) Li$_x$(C$_2$H$_8$N$_2$)$_y$MoSe$_2$ (Phase I), (c) Li$_x$(C$_2$H$_8$N$_2$)$_y$MoSe$_2$ (Phase II), (d) Li$_x$(C$_6$H$_{16}$N$_2$)$_y$MoSe$_2$.

## 3. Results and Discussion
### 3.1 $Li_x(C_2H_8N_2)_yMoSe_2$

Figure 2(a) shows the powder x-ray diffraction patterns of the host sample of 2H-MoSe$_2$ and intercalated samples of Li$_x$(C$_2$H$_8$N$_2$)$_y$MoSe$_2$ ($x$ = 0.25 − 1). The broad peak around $2\theta$ = 20° is due to the airtight sample holder. The powder x-ray diffraction patterns of the host sample of 2H-MoSe$_2$ and Li- and EDA-intercalated Li$_x$(C$_2$H$_8$N$_2$)$_y$MoSe$_2$ ($x$ = 0.25 − 1) are well indexed on the basis of the 2H-type structure ($P6_3/mmc$). The lattice parameters are estimated to be $a$ = 3.289(1) Å, $c$ = 12.92(3) Å for 2H-MoSe$_2$, which are almost the same as those in the previous reports.[3,6] For the Li- and EDA-intercalated Li$_x$(C$_2$H$_8$N$_2$)$_y$MoSe$_2$ ($x$ = 0.25), new Bragg peaks are observed at $2\theta$ ~ 7.5 and 15°, although Bragg peaks remain due to non-intercalated regions of 2H-MoSe$_2$. The $c$-axis length of Li$_x$(C$_2$H$_8$N$_2$)$_y$MoSe$_2$ ($x$ = 0.25) is estimated to be 23.7(2) Å. Thus, the interlayer spacing is expanded by ~ 5.4 Å through the co-intercalation of Li and EDA. Since the width and length of EDA are approximately 3.7 and 5.1 Å, respectively, it appears that EDA is intercalated between MoSe$_2$ layers, where the longitudinal direction of EDA is perpendicular to the MoSe$_2$ layers, as shown in Fig. 1(b). Here, we call this phase Phase I of Li$_x$(C$_2$H$_8$N$_2$)$_y$MoSe$_2$. For Li$_x$(C$_2$H$_8$N$_2$)$_y$MoSe$_2$ ($x$ = 0.5, 1), new Bragg peaks are observed at $2\theta$ ~ 7.5, 8.5° and 15, 17°, although Bragg peaks still remain due to non-intercalated regions of 2H-MoSe$_2$. The new Bragg peaks at 7.5 and 8.5° and at 15, 17° are regarded as originating from Phase I with a $c$-axis length of 23.6(8) Å and a different phase (we call this phase Phase II) with a $c$-axis length of 20.56(3) Å of Li$_x$(C$_2$H$_8$N$_2$)$_y$MoSe$_2$, respectively, as listed in Table I. In the case of Phase II, on the other hand, the expansion of the interlayer spacing is ~ 3.8 Å. Thus, EDA is speculated to be intercalated between MoSe$_2$ layers, where the longitudinal direction of EDA is parallel to the MoSe$_2$ layers, as shown in Fig. 1(c). Recently, the synthesis of Na- and EDA-intercalated Na$_x$(C$_2$H$_8$N$_2$)$_y$MoS$_2$ with a very similar crystal structure has been reported by Liyanage and Lerner.[14] According to their report, Phase I including Na instead of Li is a metastable phase that is converted to thermodynamically stable Phase II owing to the diffusion of Na and EDA. Therefore, the appearance of Phase II with increasing $x$ in Li$_x$(C$_2$H$_8$N$_2$)$_y$MoSe$_2$ may be due to the acceleration of the diffusion caused by the increase in the content of Li and EDA.

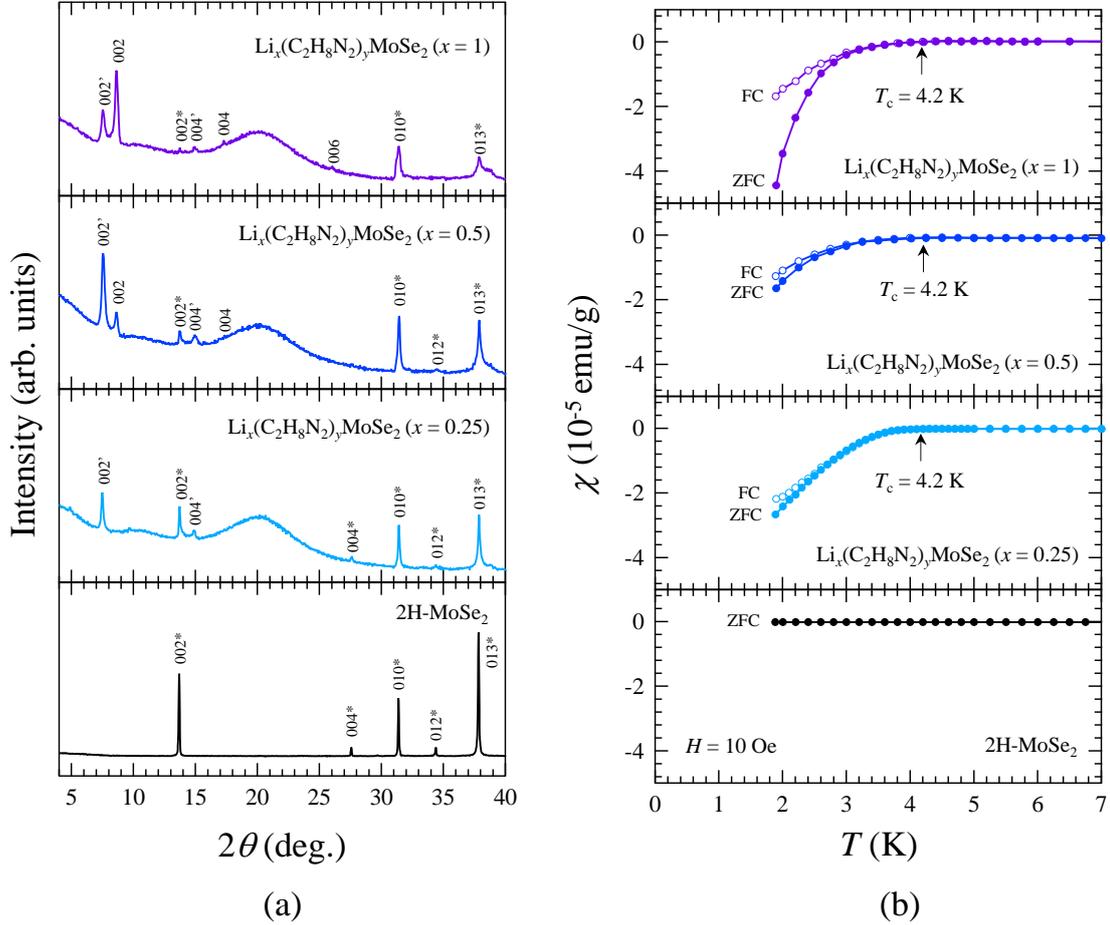

Fig. 2. (a) Powder x-ray diffraction patterns of the host sample of 2H-MoSe$_2$ and intercalated samples of Li$_x$(C$_2$H$_8$N$_2$)$_y$MoSe$_2$ ($x$ = 0.25 − 1) obtained using CuK$_\alpha$ radiation. Indices with an asterisk are due to 2H-MoSe$_2$. Indices with an apostrophe and no mark represent Phases I and II of Li$_x$(C$_2$H$_8$N$_2$)$_y$MoSe$_2$, respectively. All indices are based on the 2H-type structure ($P6_3/mmc$). The broad peak around $2\theta = 20°$ is due to the airtight sample holder. (b) Temperature dependences of the magnetic susceptibility $\chi$ in a magnetic field of 10 Oe upon zero-field cooling (ZFC) (closed circles) and field cooling (FC) (open circles) for the host sample of 2H-MoSe$_2$ and Li- and EDA- intercalated samples of Li$_x$(C$_2$H$_8$N$_2$)$_y$MoSe$_2$ ($x$ = 0.25 − 1). Arrows indicate onset temperatures of $T_c$, where $\chi$ starts to deviate from the normal-state value.

Table I. $c$-axis lengths and $T_c$ values of the host compound 2H-MoSe$_2$, as-intercalated samples, and post-annealed (250 °C, 20 h in vacuum) samples. Here, the Li content $x$ is the nominal value. $T_c$ values obtained in the magnetic susceptibility ($\chi$ - $T$) and electrical resistivity ($\rho$ - $T$) measurements are defined as the onset temperatures where $\chi$ and $\rho$ start to deviate from the normal-state values.

|  | Li content ($x$) | $c$ (Å) Phase I | $c$ (Å) Phase II | $T_c$ (K) ($\chi$ - $T$) | $T_c$ (K) ($\rho$ - $T$) | $\chi_{300K}$ ($10^{-6}$ emu/g) |
|---|---|---|---|---|---|---|
| 2H-MoSe$_2$ (host) | 0 | 12.92(3) |  | - | - | -11.53 |
| Li$_x$(C$_2$H$_8$N$_2$)$_y$MoSe$_2$ | 0.25 | 23.7(2) |  | 4.2 | 4.0 | -0.97 |
|  | 0.5 | 23.7(1) | 20.49(4) | 4.2 | 4.0 | -0.94 |
|  | 1 | 23.6(8) | 20.56(3) | 4.2 | 3.9 | -0.73 |
| Li$_x$(C$_6$H$_{16}$N$_2$)$_y$MoSe$_2$ | 0.25 | 30.8(1) |  | 3.8 |  | -1.60 |
|  | 0.5 | 37.4(4) |  | 4.2 |  | -1.11 |
|  | 1 | 37.5(4) |  | 6.0 |  | -0.18 |
| Li$_x$(C$_2$H$_8$N$_2$)$_y$MoSe$_2$ (post-annealed) → Li$_x$MoSe$_2$ | 0.25 | 13.2(1) |  | - |  | -0.88 |

Figure 2(b) shows the temperature dependences of $\chi$ in a magnetic field of 10 Oe upon zero-field cooling (ZFC) and field cooling (FC) for the host sample of 2H-MoSe$_2$ and Li- and EDA-intercalated samples of Li$_x$(C$_2$H$_8$N$_2$)$_y$MoSe$_2$ ($x = 0.25 - 1$). Although the host sample of 2H-MoSe$_2$ is non-superconducting above 1.8 K, superconducting transitions are observed at 4.2 K in the Li- and EDA-intercalated samples of Li$_x$(C$_2$H$_8$N$_2$)$_y$MoSe$_2$ ($x = 0.25 - 1$). Here, $T_c$ is defined as the onset temperature where $\chi$ starts to deviate from the normal-state value. The $T_c$ value of Li$_x$(C$_2$H$_8$N$_2$)$_y$MoSe$_2$ ($x = 0.25 - 1$) is not dependent on $x$. This is inferred to be due to the inhomogeneity of the distribution of Li$^+$ ions in the samples. That is, it appears that regions remain where Li$^+$ ions are not intercalated, and therefore the actual $x$ value in the Li- and EDA-intercalated regions is not dependent on the nominal $x$ value and is estimated to be larger than the nominal $x$ value. On the other hand, the superconducting volume fractions of these samples estimated from the $\chi$ values at 1.8 K upon ZFC are very small and less than 1%. One reason for this is that $\chi$ does not decrease completely at the measured lowest temperature of 1.8 K. Other reasons may be the inclusion of non-superconducting non-intercalated regions of 2H-MoSe$_2$ in each sample and the use of powder samples, for which the effect of the penetration depth is not negligible. In the case of Na- and NH$_3$-intercalated Na$_x$(NH$_3$)$_y$MoSe$_2$ with $T_c = 6$ K, it has been reported that the superconducting volume fraction estimated from $\chi$ measurements is much smaller in a polycrystalline powder sample than in the single crystal.[6] For the single crystal of Na$_x$(NH$_3$)$_y$MoSe$_2$, the superconducting volume fraction reaches 100% at 2.5 K. On the other hand, it has also been reported that the superconducting volume fraction of a polycrystalline powder sample of $Na_x(NH_3)_yMoSe_2$ is very small and less than 1% at 2.5 K. In the case that the size of superconducting grains is comparable to the penetration depth, a small superconducting volume fraction is often observed in powder samples.[6,9,15,16] Since a polycrystalline powder sample of $2H$-$MoSe_2$ was used for the host sample, it appears that the superconducting volume fraction was estimated from the $\chi$ values to be less than 1%. Therefore, it is concluded that Li- and EDA-intercalated regions are actually a superconducting state.

The superconductivity observed in the $\chi$ measurements for the intercalated samples has been confirmed in the $\rho$ measurements. Figures 3(a) and 3(b) show the temperature dependences of $\rho$ for the host sample of $2H$-$MoSe_2$ and the intercalated samples of $Li_x(C_2H_8N_2)_yMoSe_2$ ($x = 0.25 - 1$). It has been found that the host sample of $2H$-$MoSe_2$ exhibits semiconducting behavior of $\rho$. For the intercalated samples of $Li_x(C_2H_8N_2)_yMoSe_2$ ($x = 0.25 - 1$), the value of $\rho$ decreases to $10^{-1} - 10^{-2}$ $\Omega\cdot$cm. This indicates that the intercalated samples are metallic, although the temperature dependence of $\rho$ is not so metallic due to insulating grain boundaries.

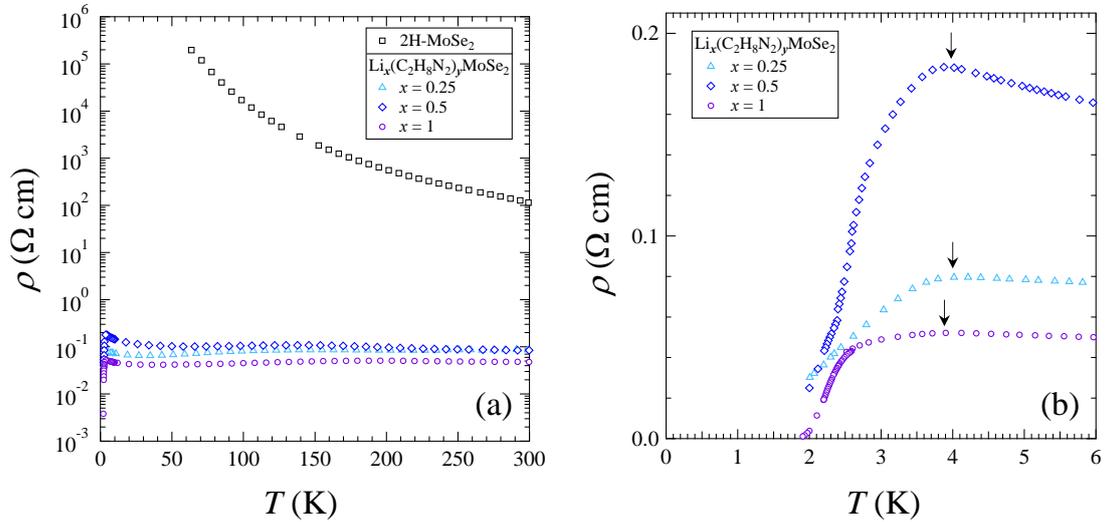

Fig. 3. (a) Temperature dependences of the electrical resistivity $\rho$ for the host sample of $2H$-$MoSe_2$ and Li- and EDA-intercalated samples of $Li_x(C_2H_8N_2)_yMoSe_2$ ($x = 0.25 - 1$), which were pelletized at room temperature without heat treatment. (b) Enlarged view of (a) at low temperatures around $T_c$. Arrows indicate onset temperatures of $T_c$, where $\rho$ starts to deviate from the normal-state value.

As shown in Fig. 3(b), all the intercalated samples show a superconducting transition at a low temperature below 4 K. Values of $T_c$, defined as the onset temperature where $\rho$ starts to deviate from the normal-state value, are listed in Table I. These results indicate that the electron doping due to the Li intercalation induces a metallic state, leading to the appearance of superconductivity. Note that the $T_c$ values obtained in the $\rho$ measurements are slightly lower than those obtained in the $\chi$ measurements, as listed in Table I. In addition, the superconducting transitions are very broad and the zero resistivity is not observed above 2 K except for $x = 1$. These findings are inferred to be due to insulating grain boundaries in the intercalated pellet samples and also due to the degradation of the samples caused by the atmospheric exposure in the process of forming four terminals on the sample surface for the $\rho$ measurements.

### 3.2 $Li_x(C_6H_{16}N_2)_yMoSe_2$

Figure 4(a) shows the powder x-ray diffraction patterns of the host sample of 2H-MoSe$_2$ and Li- and HMDA-intercalated samples of $Li_x(C_6H_{16}N_2)_yMoSe_2$ ($x = 0.25 - 1$). New Bragg peaks are observed owing to the intercalation, although Bragg peaks due to non-intercalated regions of 2H-MoSe$_2$ also remain. The powder x-ray diffraction patterns of $Li_x(C_6H_{16}N_2)_yMoSe_2$ ($x = 0.25 - 1$) are well indexed on the basis of the 2H-type structure ($P6_3/mmc$). For $Li_x(C_6H_{16}N_2)_yMoSe_2$ ($x = 0.25$), new Bragg peaks are observed at $2\theta \sim 5.5$ and 11°. Thus, the $c$-axis length is estimated to be 30.8(1) Å and the interlayer spacing is expanded by $\sim 8.9$ Å through the co-intercalation of Li and HMDA. Since the width and length of HMDA are approximately 3.7 and 10.4 Å, respectively, HMDA is speculated to be intercalated between the MoSe$_2$ layers, where the longitudinal direction of HMDA is oblique to the MoSe$_2$ layers, as shown in Fig. 1(d). Such an obliquely intercalated structure has also been speculated to exist in Li- and HMDA-intercalated $Li_x(C_6H_{16}N_2)_yTiSe_2$[9] and $Li_x(C_6H_{16}N_2)_yFe_{2-z}Se_2$.[17,18] For $Li_x(C_6H_{16}N_2)_yMoSe_2$ ($x = 0.5, 1$), new Bragg peaks are observed at $2\theta \sim 4.5$ and 9°. Thus, the $c$-axis lengths of $Li_x(C_6H_{16}N_2)_yMoSe_2$ ($x = 0.5, 1$) are estimated to be 37.4(4) and 37.5(4) Å, respectively, and the interlayer spacings are expanded by $\sim 12.3$ Å through the co-intercalation of Li and HMDA. Taking into account the size of HMDA, the state of HMDA between the MoSe$_2$ layers in $Li_x(C_6H_{16}N_2)_yMoSe_2$ ($x = 0.5, 1$) has not been determined definitely.

The temperature dependences of $\chi$ in a magnetic field of 10 Oe upon ZFC and FC for Li- and HMDA-intercalated $Li_x(C_6H_{16}N_2)_yMoSe_2$ ($x = 0.25 - 1$) are shown in Fig. 4(b). It is found that the co-intercalated samples of $Li_x(C_6H_{16}N_2)_yMoSe_2$ ($x = 0.25 - 1$) show a superconducting transition at 3.8 − 6.0 K and that the $T_c$ values tend to increase with increasing *x*. This result is in contrast to that of Li- and EDA-intercalated Li$_x$(C$_2$H$_8$N$_2$)$_y$MoSe$_2$ ($x = 0.25 - 1$), in which the $T_c$ values are independent of *x*. The superconducting volume fractions of these samples estimated from the change in $\chi$ below $T_c$ are as small as those in the case of Li$_x$(C$_2$H$_8$N$_2$)$_y$MoSe$_2$.

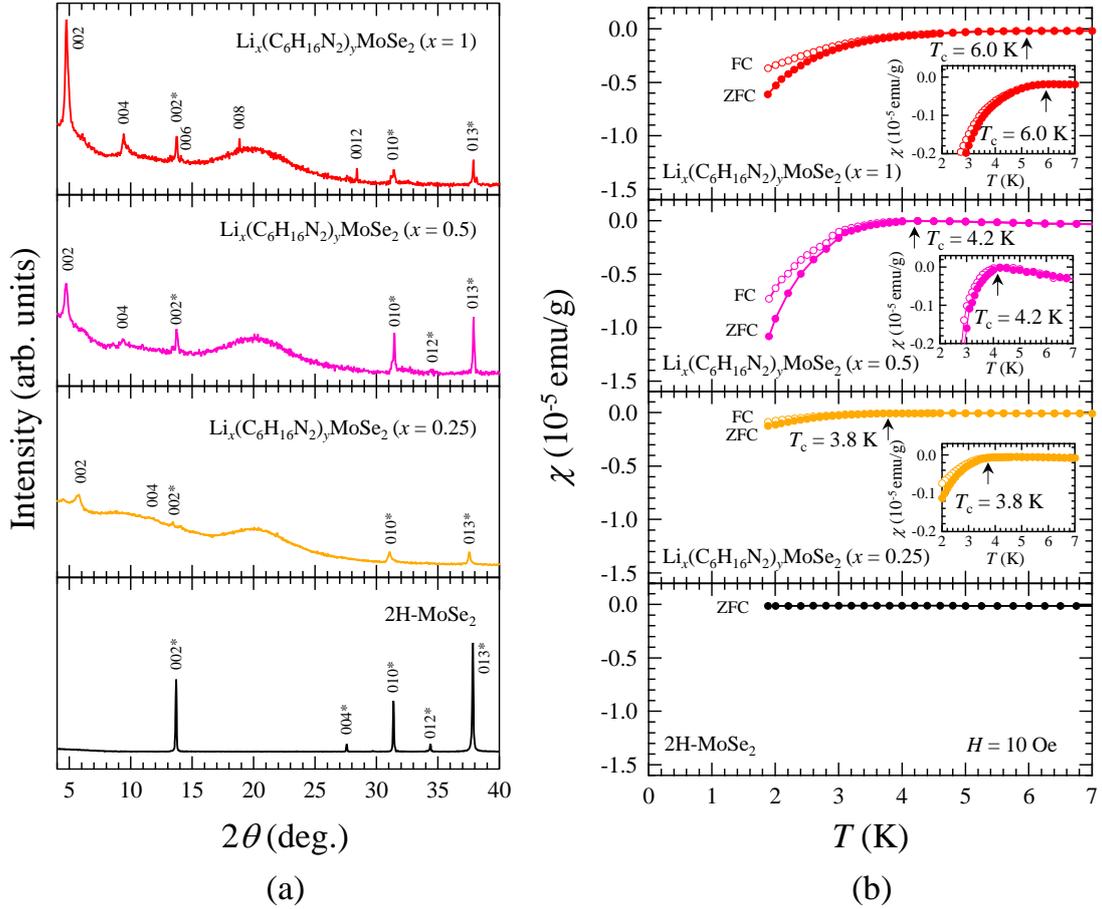

Fig. 4. (a) Powder x-ray diffraction patterns of the host sample of 2H-MoSe$_2$ and Li- and HMDA-intercalated samples of Li$_x$(C$_6$H$_{16}$N$_2$)$_y$MoSe$_2$ ($x = 0.25 - 1$) obtained using CuK$_\alpha$ radiation. Indices with and without an asterisk are due to 2H-MoSe$_2$ and Li$_x$(C$_6$H$_{16}$N$_2$)$_y$MoSe$_2$, respectively. All indices are based on the 2H-type structure ($P6_3/mmc$). The broad peak around $2\theta = 20°$ is due to the airtight sample holder. (b) Temperature dependences of the magnetic susceptibility $\chi$ in a magnetic field of 10 Oe upon zero-field cooling (ZFC) (closed circles) and field cooling (FC) (open circles) for the host sample of 2H-MoSe$_2$ and Li- and HMDA-intercalated samples of Li$_x$(C$_6$H$_{16}$N$_2$)$_y$MoSe$_2$ ($x = 0.25 - 1$). Arrows indicate onset temperatures of $T_c$, where $\chi$ starts to deviate from the normal-state value.

## 3.3 Post-annealing effects of $Li_x(C_2H_8N_2)_yMoSe_2$

For the deintercalation of EDA, TG measurements were carried out. Figure 5 shows the TG curves obtained upon heating up to 800°C at a rate of 1 °C/min for the host sample of 2H-MoSe$_2$ and Li- and EDA-intercalated samples of $Li_x(C_2H_8N_2)_yMoSe_2$ ($x$ = 0.25, 0.5). In these samples, the increase in mass with increasing temperature above ~ 500°C may be attributed to the formation of MoO$_3$. For the intercalated samples, several steps of mass loss are observed. It is inferred that the first loss below ~ 100°C is due to the desorption of EDA on the surface of grains and that the second loss between ~ 100°C and 500°C is due to the deintercalation of EDA. The EDA contents $y$ in the as-intercalated samples of $Li_x(C_2H_8N_2)_yMoSe_2$ ($x$ = 0.25, 0.5) are estimated from the second loss to be ~ 0.09 and 0.24, respectively. It is found that the $y$ value increases with increasing $x$ and is roughly half of the $x$ value. This is reasonable because Li$^+$ ions seem to be located near lone-pair electrons of N atoms at both edges of linear molecules of EDA, as shown in Fig. 1.

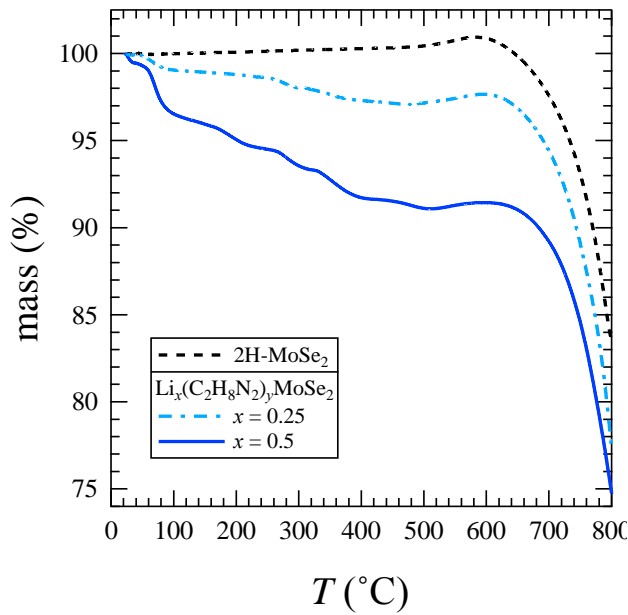

Fig. 5. Thermogravimetric curves obtained upon heating at a rate of 1 °C/min for the host sample of 2H-MoSe$_2$ and Li- and EDA-intercalated samples of $Li_x(C_2H_8N_2)_yMoSe_2$ ($x$ = 0.25, 0.5).

Figure 6(a) shows the powder x-ray diffraction pattern of $Li_x(C_2H_8N_2)_yMoSe_2$ ($x$ = 0.25) post-annealed in vacuum at 250°C for 20 h. Those of the as-intercalated sample of $Li_x(C_2H_8N_2)_yMoSe_2$ ($x$ = 0.25) and the host sample of 2H-MoSe$_2$ are also shown for reference. It is found that the post-annealed sample of $Li_x(C_2H_8N_2)_yMoSe_2$ ($x$ = 0.25) changes to Li$_x$MoSe$_2$ ($x$ = 0.25) and 2H-MoSe$_2$ from Phase I of $Li_x(C_2H_8N_2)_yMoSe_2$ ($x$ = 0.25) due to the deintercalation of EDA, because Li$^+$ ions are known not to deintercalate at a temperature as low as 250°C.[9,18] In fact, the $c$-axis length of Li$_x$MoSe$_2$ ($x$ = 0.25) is 13.2(1) Å and slightly longer than that of 2H-MoSe$_2$, as listed in Table I. Also in the case of Li- and EDA-intercalated $Li_x(C_2H_8N_2)_yTiSe_2$, $Li_xTiSe_2$ is formed through the deintercalation of EDA by post-annealing in vacuum at 250°C.[9]

Figure 6(b) shows the temperature dependences of $\chi$ in a magnetic field of 10 Oe upon ZFC for the post-annealed and as-intercalated samples of $Li_x(C_2H_8N_2)_yMoSe_2$ ($x = 0.25$). Although the as-intercalated sample of $Li_x(C_2H_8N_2)_yMoSe_2$ ($x = 0.25$) shows a superconducting transition at 4.2 K, it is found that the post-annealed sample is non-superconducting above 1.8 K. This indicates that $Li_xMoSe_2$ ($x = 0.25$) is not superconducting and that both Li- and EDA-intercalated $Li_x(C_2H_8N_2)_yMoSe_2$ is a new superconductor with $T_c = 4.2$ K.

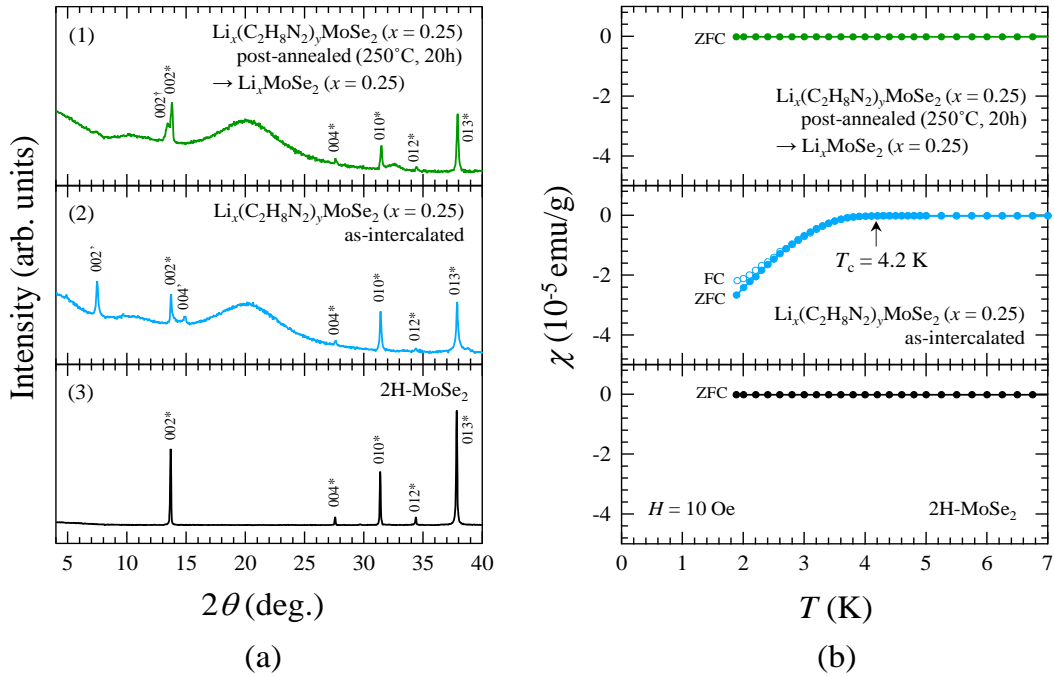

Fig. 6. (a) Powder x-ray diffraction pattern of $Li_x(C_2H_8N_2)_yMoSe_2$ ($x = 0.25$) post-annealed in vacuum at 250°C for 20 h obtained using CuK$_\alpha$ radiation. Those of the host sample of 2H-MoSe$_2$ and the as-intercalated sample before the post-annealing are also shown for reference. Indices with an asterisk, an apostrophe, and a dagger represent 2H-MoSe$_2$, Phase I of $Li_x(C_2H_8N_2)_yMoSe_2$, and $Li_xMoSe_2$, respectively. All indices are based on the 2H-type structure ($P6_3/mmc$). The broad peak around $2\theta = 20°$ is due to the airtight sample holder. (b) Temperature dependence of the magnetic susceptibility $\chi$ in a magnetic field of 10 Oe upon zero-field cooling (ZFC) for $Li_x(C_2H_8N_2)_yMoSe_2$ ($x = 0.25$) post-annealed in vacuum at 250°C for 20 h. Those of the host sample of 2H-MoSe$_2$ and the as-intercalated sample before the post-annealing are also shown for reference.

### 3.4 Pauli paramagnetism and relationship with $T_c$

Figure 7 shows the temperature dependences of $\chi$ in a magnetic field of 1 T upon ZFC for the host sample of 2H-MoSe$_2$, as-intercalated samples of Li$_x$(C$_2$H$_8$N$_2$)$_y$MoSe$_2$ ($x$ = 0.25 − 1) and Li$_x$(C$_6$H$_{16}$N$_2$)$_y$MoSe$_2$ ($x$ = 0.25 − 1), and the sample of Li$_x$MoSe$_2$ ($x$ = 0.25) obtained by the post-annealing of Li$_x$(C$_2$H$_8$N$_2$)$_y$MoSe$_2$ ($x$ = 0.25) in vacuum at 250°C for 20 h. It is found that the $\chi$ values of all the intercalated samples are much larger than those of the host sample of 2H-MoSe$_2$ and almost independent of the temperature except for at low temperatures, indicating the increase in the Pauli paramagnetism in the intercalated samples. Since the Pauli paramagnetism is proportional to the EDOS at the Fermi level, this suggests that intercalated Li supplies MoSe$_2$ layers with electron carriers, leading to the increase in the Pauli paramagnetism, namely, the increase in the EDOS at the Fermi level. Note that all the intercalated samples contain non-intercalated regions of 2H-MoSe$_2$. Therefore, it is inappropriate to compare the $\chi$ values of the intercalated samples owing to the inhomogeneous distribution of Li$^+$ ions in a sample. However, it is clear that intercalated regions induce larger $\chi$ values than non-intercalated regions of 2H-MoSe$_2$. Moreover, it is also found that the $\chi$ value increases through the deintercalation of EDA by the post-annealing. This may be due to the possible transfer to MoSe$_2$ layers of valence electrons of intercalated Li, which were partly used for the bonding with EDA, leading to the increase in the Pauli paramagnetism. Such an increase in the Pauli paramagnetism due to the deintercalation of EDA is also observed in Li- and EDA-intercalated Li$_x$(C$_2$H$_8$N$_2$)$_y$TiSe$_2$.[9]

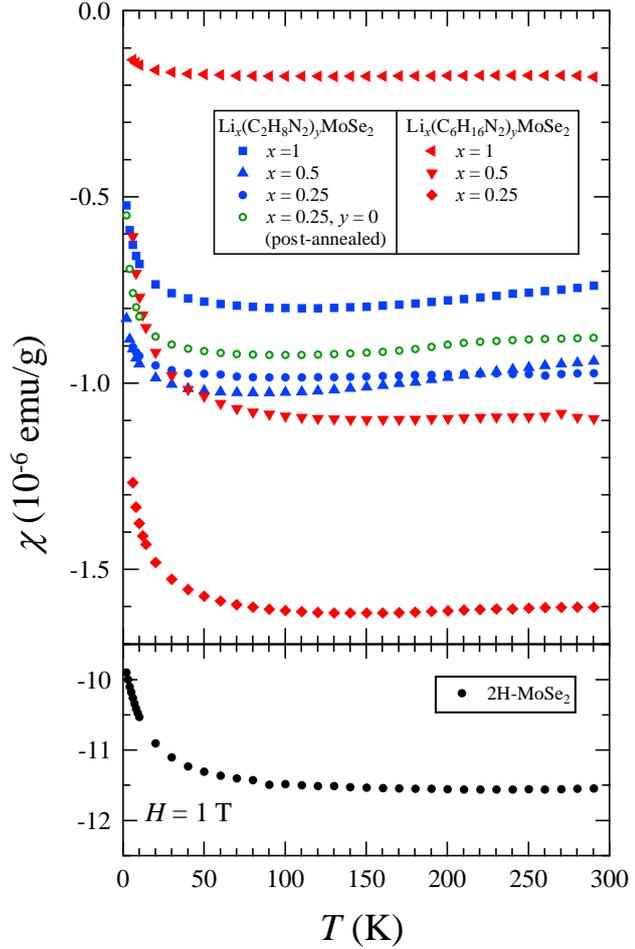

Fig. 7. Temperature dependences of the magnetic susceptibility $\chi$ in a magnetic field of 1 T upon zero-field cooling for the host sample of 2H-MoSe$_2$, as-intercalated samples of Li$_x$(C$_2$H$_8$N$_2$)$_y$MoSe$_2$ ($x$ = 0.25 − 1) and Li$_x$(C$_6$H$_{16}$N$_2$)$_y$MoSe$_2$ ($x$ = 0.25 − 1), and the sample of Li$_x$MoSe$_2$ ($x$ = 0.25) obtained by the post-annealing of Li$_x$(C$_2$H$_8$N$_2$)$_y$MoSe$_2$ ($x$ = 0.25) in vacuum at 250°C for 20 h.

Finally, we discuss the relationship between $T_c$ and the interlayer spacing between MoSe$_2$ layers or $\chi$ at 300 K, $\chi_{300K}$, for intercalated samples of Li$_x$(C$_2$H$_8$N$_2$)$_y$MoSe$_2$ ($x$ = 0.25 − 1) and Li$_x$(C$_6$H$_{16}$N$_2$)$_y$MoSe$_2$ ($x$ = 0.25 − 1), as shown in Figs. 8(a) and 8(b).

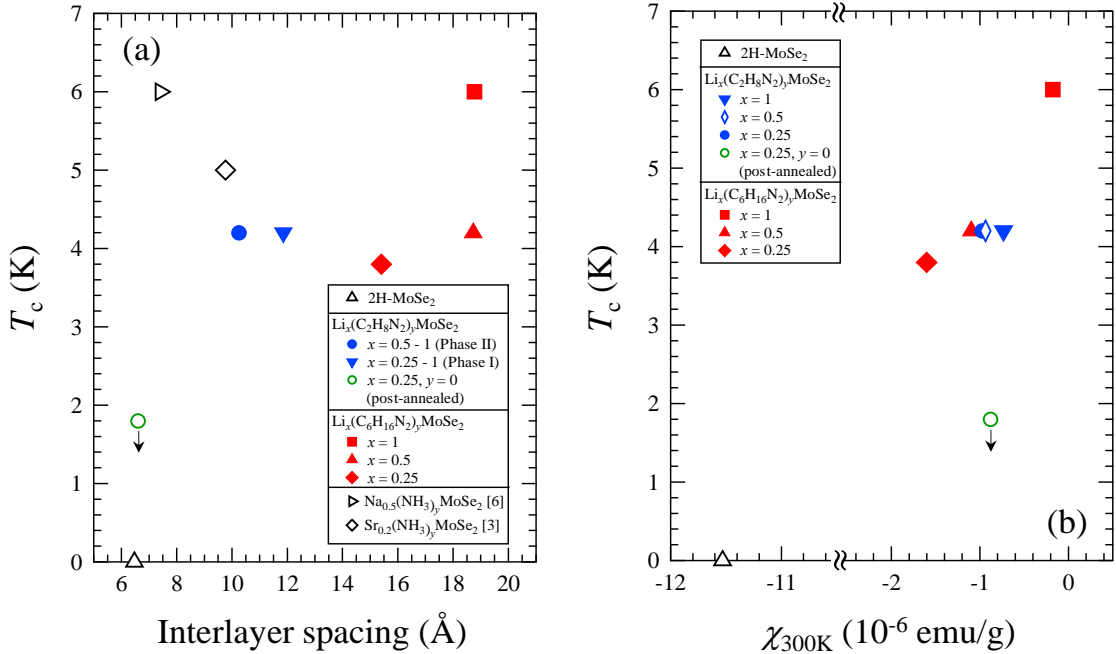

Fig. 8. Relationship between $T_c$ and (a) the interlayer spacing between MoSe$_2$ layers or (b) $\chi$ at 300 K, $\chi_{300K}$, for the host sample of 2H-MoSe$_2$, as-intercalated samples of Li$_x$(C$_2$H$_8$N$_2$)$_y$MoSe$_2$ ($x$ = 0.25 − 1) and Li$_x$(C$_6$H$_{16}$N$_2$)$_y$MoSe$_2$ ($x$ = 0.25 − 1), and the sample of Li$_x$MoSe$_2$ ($x$ = 0.25) obtained by the post-annealing of Li$_x$(C$_2$H$_8$N$_2$)$_y$MoSe$_2$ ($x$ = 0.25) in vacuum at 250°C for 20 h. Arrows indicate samples that are not superconducting above 1.8 K. The data of $A_x$(NH$_3$)$_y$MoSe$_2$ ($A$ = Na, Sr) in former papers[3,6] are also shown in (a) for reference.

It is found that there is no clear correlation between $T_c$ and the interlayer spacing. This is obvious, taking into account the data of $A_x$(NH$_3$)$_y$MoSe$_2$ ($A$ = Na, Sr) in former papers[3,6] shown in Fig. 8(a). In the case of the co-intercalated samples, it is possible that the electronic structure is perfectly two-dimensional owing to the inclusion of EDA,

HMDA, or $NH_3$ between $MoSe_2$ layers. Thus, the $T_c$ value is not dependent on the interlayer spacing. On the other hand, the $T_c$ values are correlated with $\chi_{300K}$. The $\chi_{300K}$ values of the intercalated samples are much larger than that of the host sample of 2H-$MoSe_2$. Therefore, it follows that the increase in $\chi_{300K}$, namely, the increase in the EDOS at the Fermi level is necessary for the appearance of superconductivity in $MoSe_2$-based intercalation compounds. Figure 8(b) reveals that the $T_c$ values tend to increase with increasing $\chi_{300K}$, but this is not always the case due to the inclusion of non-intercalated regions of 2H-$MoSe_2$ in the intercalated samples. Here, it is noteworthy that only Li-intercalated $Li_xMoSe_2$ ($x = 0.25$) with a small interlayer spacing is non-superconducting, even though its $\chi_{300K}$ value is comparable to those of Li- and EDA-intercalated $Li_x(C_2H_8N_2)_yMoSe_2$ and Li- and HMDA-intercalated $Li_x(C_6H_{16}N_2)_yMoSe_2$. Accordingly, it is concluded that not only a sufficient amount of EDOS at the Fermi level due to the charge transfer from intercalated Li to $MoSe_2$ layers but also the enhancement of the two-dimensionality of the crystal structure and/or electronic structure due to the expansion of the interlayer spacing between $MoSe_2$ layers is necessary for the appearance of superconductivity in $MoSe_2$-based intercalation compounds. The former is reasonable because 2H-$MoSe_2$ is semiconducting and short of carriers for superconductivity. Regarding the latter, in fact, there has been no report on the appearance of superconductivity in only alkali metal- or alkaline-earth metal-intercalated $A_xMoSe_2$, although both $A$- and $NH_3$-intercalated $A_x(NH_3)_yMoSe_2$ ($A$ = Li, Na, K, Sr) with a large interlayer spacing is superconducting as mentioned in sect. 1.[3,6] Moreover, the superconductivity in the surface layer of thin flakes of 2H-$MoX_2$ ($X$ = S, Se, Te) induced by the EDLT technique[7,8] suggests the importance of the two-dimensionality in the appearance of superconductivity in $MoSe_2$-based compounds. Accordingly, it is possible that the enhancement of the electron-phonon interaction in each $MoSe_2$ layer and/or the electronic polarization of EDA or HMDA contribute to the formation of Cooper pairs. It is also speculated that the pairing mechanism in the $MoSe_2$-based intercalation superconductors with large interlayer spacings is similar to that in an EDLT of 2H-$MoX_2$ ($X$ = S, Se, Te). Here, we comment on the increase in $T_c$ with increasing $c$-axis length owing to the increase in the ionic radius of $A$ in $A_x(NH_3)_yMoSe_2$ ($A$ = Li, Na, K, Sr) pointed by Miao et al.[6] and also in $A_x(NH_3)_yMoS_2$ ($A$ = Li, Na, K, Rb, Cs, Ca, Sr, Ba, Yb).[3,4] Their insistence may be correct within a limited range of samples where only $A$ varies. As shown in Fig. 8(a), $T_c$ values have no correlation with the interlayer spacing, namely, the $c$-axis length among various $MoSe_2$-based intercalation superconductors including the present samples. The $T_c$ values of $A_x(NH_3)_yMoSe_2$ are expected to depend on the EDOS at the Fermi level, as shown in Fig. 8(b), because an increase in the $c$-axis length simply leads to an increase in the EDOS at the Fermi level.

## 4. Summary

We have successfully synthesized new intercalation superconductors of $Li_x(C_2H_8N_2)_yMoSe_2$ and $Li_x(C_6H_{16}N_2)_yMoSe_2$ with $T_c$ = 4.2 and 3.8 – 6.0 K, respectively, via the co-intercalation of lithium and EDA or HMDA into semiconducting 2H-MoSe$_2$. It has been found that the $T_c$ values are related not to the interlayer spacing between MoSe$_2$ layers so much but to the Pauli paramagnetism, namely, to the EDOS at the Fermi level, which is increased by the electron doping due to intercalated Li. Moreover, it has been found that only Li-intercalated $Li_xMoSe_2$ ($x$ = 0.25) prepared through the deintercalation of EDA from $Li_x(C_2H_8N_2)_yMoSe_2$ ($x$ = 0.25) with $T_c$ = 4.2 K is non-superconducting. Accordingly, it has been concluded that not only a sufficient amount of EDOS at the Fermi level due to the charge transfer from intercalated Li to MoSe$_2$ layers but also the enhancement of the two-dimensionality of the crystal structure and/or electronic structure due to the expansion of the interlayer spacing between MoSe$_2$ layers is necessary for the appearance of superconductivity in MoSe$_2$-based intercalation superconductors. It has also been concluded that the enhancement of the electron-phonon interaction in each MoSe$_2$ layer due to the increase in the interlayer spacing and/or the electronic polarization of EDA or HMDA may contribute to the formation of Cooper pairs, which may be analogous to the superconductivity in the surface layer of thin flakes of 2H-Mo$X_2$ ($X$ = S, Se, Te) induced by the EDLT technique.

**Acknowledgments**

This work was supported by JSPS KAKENHI (Grant Numbers 15K13512 and 16K05429). One of the authors (K. S.) was supported by the Tohoku University Division for Interdisciplinary Advanced Research and Education.